\documentclass[a4paper,notitlepage,11pt]{article}

\usepackage[ansinew]{inputenc}
\usepackage[T1]{fontenc}   
\usepackage{graphicx} 
\usepackage[lmargin=2cm, rmargin=2cm,bottom=2cm,top=2cm]{geometry} 
\usepackage{amsmath}
\usepackage{amsthm} 
\usepackage{amssymb}  
\usepackage{url}
\usepackage{cite}
\usepackage{dsfont}
\usepackage{enumerate}
\usepackage{multicol}
\usepackage{color}
\usepackage{caption}
\usepackage[rgb,usenames,dvipsnames]{xcolor}
\usepackage{cite}
\usepackage{xifthen}
\usepackage[rightcaption]{sidecap}  
\usepackage[boxed]{algorithm2e} \SetAlCapSkip{.2cm}
\usepackage{psfrag}
\usepackage{tikz}
\usetikzlibrary{matrix,calc,decorations.pathreplacing}
\usepackage{mathtools}

\newcommand*\xbar[1]{%
  \hbox{%
    \vbox{%
      \hrule height 0.5pt 
      \kern0.5ex
      \hbox{%
        \kern-0.1em
        \ensuremath{#1}%
        \kern-0.1em
      }%
    }%
  }%
} 

\def\p{\pi}
\newcommand{\States}{\mathcal{S}}
\newcommand{\Actions}{\mathcal{A}}

\newcommand{\val}[1][\p]{v^{#1}}

\newcommand{\T}[1][\p]{T^{#1}}
\renewcommand{\S}[1][\p]{S^{#1}}

\DeclarePairedDelimiter\floor{\lfloor}{\rfloor}
\newcommand{\nchoosek}[2]{\begin{pmatrix} #1 \\ #2 \end{pmatrix}}
\newcommand{\smallnchoosek}[2]{\bigl(\begin{smallmatrix} #1 \\ #2 \end{smallmatrix}\bigr)}
\newcommand{\ra}{\longrightarrow}

\newtheorem{thm}{Theorem}

\newtheorem{prop}{Proposition}

\newtheorem{lem}{Lemma}
\newtheorem{cor}{Corollary}
\theoremstyle{definition}
\newtheorem{defi}{Definition} 
\newtheorem{pb}{Problem}
\newtheorem{rel}{Relaxation} 

\theoremstyle{remark}

\begin{document}

\title { Improved bound on the worst case complexity of Policy Iteration }
\author{ Romain Hollanders, Balázs Gerencsér, Jean-Charles Delvenne and Raphaël M. Jungers\thanks{This work was supported by an ARC grant from the French Community of Belgium and by the IAP network 'Dysco' funded by the office of the Prime Minister of Belgium. The scientific responsiblity rests with the authors. J.-C. D. is with CORE and NAXYS. R. M. J. is an F.R.S./FNRS Research Associate.}}
\date{}
\maketitle

\setlength{\parindent}{0pt}
\setlength{\parskip}{1ex plus 0.5ex minus 0.2ex}

\begin{abstract}

Solving Markov Decision Processes (MDPs) is a recurrent task in engineering. Even though it is known that solutions for minimizing the infinite horizon expected reward can be found in polynomial time using Linear Programming techniques, iterative methods like the Policy Iteration algorithm (PI) remain usually the most efficient in practice. This method is guaranteed to converge in a finite number of steps. Unfortunately, it is known that it may require an exponential number of steps in the size of the problem to converge. On the other hand, many open questions remain considering the actual worst case complexity. In this work, we provide the first improvement over the fifteen years old upper bound from Mansour \& Singh (1999) by showing that PI requires at most $\frac{k}{k-1} \cdot \frac{k^n}{n} + o\big(\frac{k^n}{n}\big)$ iterations to converge, where $n$ is the number of states of the MDP and $k$ is the maximum number of actions per state. Perhaps more importantly, we also show that this bound is optimal for an important relaxation of the problem.

\end{abstract}

\section{Introduction}

Markov Decision Processes (MDPs) have been found to be a powerful modeling tool for the decision problems that arise daily in various domains of engineering such as control~\cite{Bertsekas2007}, finance~\cite{bauerle2011}, communication networks~\cite{altman2002}, queuing systems\cite{meyn2008}, PageRank optimization~\cite{csaji2014}, and many more (see~\cite{White1993} for a more exhaustive list).
MDPs are described from a set of $n$ states in which a system can be. When being in a state, the controller of the system must choose an available action in that state, each of which induces a reward and moves the system to another state according to given transition probabilities. In this work, we assume that the number of actions per state is bounded by a constant $k$. A policy refers to the stationary choice of one action in every state. Choosing a policy implies fixing a dynamics that corresponds to a Markov chain. Given any policy (there are at most $k^n$ of them), we can associate a value to each state of the MDP that corresponds to the infinite-horizon expected reward of an agent starting in that state. By solving an MDP, we mean providing an optimal policy that maximizes the value of every state. Depending on the application, a total-, discounted- or average-reward criterion may be best suited to define the value function. Note that in every case, an optimal policy always exists. See e.g.~\cite{Bertsekas2007} and~\cite{Puterman1994} for a comprehensive and in-depth study of MDPs.

One practically efficient way of finding the optimal policy for an MDP is to use the Policy Iteration algorithm (PI). Starting from an initial policy $\pi_0$, $i=0$, this simple iterative scheme repeatedly computes the value of $\pi_i$ at every state and greedily modifies this policy using its evaluation to obtain the next iterate $\pi_{i+1}$. The modification always ensures that the value of $\pi_{i+1}$ improves on that of $\pi_i$ at every state. The process is then repeated until convergence to the optimal policy $\pi^*$ in a finite number of steps (obviously at most $k^n$ steps---the maximum number of policies). We refer to the ordered set of explored policies as the PI-sequence. A more precise statement of the algorithm as well as some important properties are described in Section~\ref{sec:preliminaries}.

Every iteration of the algorithm can be performed in polynomial time and its number of steps has been shown to be strongly polynomial in some important particular cases such as discounted-reward MDPs with a fixed discount rate~\cite{Tseng1990,Ye2011} or deterministic MDPs~\cite{post2013} (the bounds in these results were later improved in \cite{hansen2013} and \cite{scherrer2013}). However, in the general case the number of iterations of PI can be exponentially large. Based on the work of Friedmann on Parity Games~\cite{friedmann2009}, PI has been shown to require at least $\Omega(2^{n/7})$ steps to converge in the worst case for the total- and average-reward criteria \cite{Fearnley2010} and for the discounted-reward criterion~\cite{hollanders2012}. Friedmann's result was also a major milestone for the study of the Simplex algorithm for Linear Programming as it lead to exponential lower bounds for some critical pivoting rules~\cite{friedmann2011B,friedmann2011C}. On the other hand, the best known upper bound for PI to date was due to Mansour \& Singh with a $13\cdot\frac{k^n}{n}$ steps bound \cite{Mansour1999}. In Section~\ref{sec:upper-bound}, we provide the first improvement in fifteen years over Mansour \& Singh's bound, namely $\frac{k}{k-1} \cdot \frac{k^n}{n} + o\big(\frac{k^n}{n}\big)$. 

PI-sequences need to verify several combinatorial conditions that have been identified over the years \cite{Mansour1999,madani1998,szabo2001,holt1999}. However to obtain our bound, we only exploit a subset of these conditions.
In Section~\ref{sec:relaxation}, we define the notion of Pseudo-PI-sequence that describes any sequence of policies satisfying this subset of conditions. We then prove in Section~\ref{sec:thm1} that the above upper bound holds for both PI- and Pseudo-PI-sequences. As it turns out, our bound is tight for Pseudo-PI-sequences. Indeed, in Section~\ref{sec:lower-bound} we provide a construction of a Pseudo-PI-sequence of length $\frac{k}{k-1} \cdot \frac{k^n}{n} + o\big(\frac{k^n}{n}\big)$. We believe that this construction is important in that it shows that additional properties of PI-Sequences must be exploited if a tighter bound is to be obtained, see Sect~\ref{sec:perspectives}.


\section{Problem statement} \label{sec:preliminaries}


\begin{defi}[Markov Decision Process]
	Let $\States = \{1, \dots, n\}$ be a set of $n$ \emph{states} and $\Actions_s$
	be a set of $k$ \emph{actions} available for state $s \in \States$. To each choice of these actions corresponds a \emph{transition probability} distribution for the next state to visit as well as a \emph{reward}. For simplicity, we use a common numbering for the actions, that is, $\Actions_s \triangleq \Actions = \{1, \dots, k\}$ for all $s \in \States$. With this notation, for every pair $(s, a) \in \States \times \Actions$, the transition probability and reward functions are uniquely defined. Let us call a \emph{policy} $\p \in \{ 1, \dots, k \}^n$ the stationary choice of one action for every state. Every policy induces a given transition probability matrix $P^{\p}$ corresponding to some Markov chain and a reward vector $r^{\p}$. We may ask how rewarding a policy $\p$ is in the long run. This is represented by its \emph{value} vector $\val \in \mathbb{R}^n$ whose $s^{\text{th}}$ entry corresponds to the long term reward obtained from starting in state $s$ and following the policy $\pi$ thereafter. It can be computed by solving a system whose definition depends on the problem studied. For instance for the standard \emph{infinite-horizon average reward criterion} where the aim is to maximize the average reward at each step, $\val$ is obtained by:
	\begin{align*}
		\val = \limsup_{N \rightarrow \infty} \frac{1}{N} \sum_{i=0}^{N-1} \left( P^{\p} \right)^i \ r^{\p}.
	\end{align*} 
	However, in this work, the bounds that we derive do not depend on the chosen reward criterion. By \emph{solving} an MDP, we mean finding the optimal policy $\p^*$ such that for any other policy $\p$, $\val[\p^*] \geq \val$, that is, $\val[\p^*](s) \geq \val(s)$ for all states $s$. The existence of such a policy is guaranteed~\cite{Bertsekas2007}.
\end{defi}

\begin{defi}[Domination]
	Given two policies $\p$ and $\p'$, if $\val[\p'](s) \geq \val(s)$ for all states $s \in \States$, we say that $\p'$ \emph{dominates} $\p$ and we write $\p' \succeq \p$. If moreover $\val[\p'](s) > \val(s)$ for at least one state, then the domination is strict and we write $\p' \succ \p$.
\end{defi}

\begin{defi}[Switching]
	Let $U$ be a collection of state-action pairs $(s, a)$. We say that $U$ is \emph{well-defined} if it contains every state $s \in \States$ at most once. In that case, we define $\p' = \p \oplus U$ to be the policy obtained from $\p$ by \emph{switching} the action $\p(s)$ to $a$ for each $(s, a)$-pair in $U$.
\end{defi}

\begin{defi}[Improvement set]
	We define the \emph{improvement set} of a policy $\p$ as:
	\begin{align*}
		\T = \big\{(s,a) \ | \ \p \oplus \{(s,a)\} \succ \p \big\},
	\end{align*}
	and the set of \emph{improvement states} $\S$ of $\p$ as the set of states that appear in $\T$.
\end{defi}

\begin{prop}[Proposition 1.3.4 in \cite{Bertsekas2007}, Volume 2] \label{thm:MS1}
	Let $\p$ be a policy and $U \neq \emptyset$ be any well-defined subset of its improvement set $\T$. Then $\p \oplus U \succ \p$.
\end{prop}

\begin{prop}[Proposition 1.3.4 in \cite{Bertsekas2007}, Volume 2] \label{thm:MS2}
	For a given policy $\p$, if $\T = \emptyset$, then $\p$ is optimal.
\end{prop}

Based on Propositions~\ref{thm:MS1} and~\ref{thm:MS2} we may define the \emph{Policy Iteration} algorithm to find the optimal policy.

\begin{algorithm}[h] 
	\SetAlgoLined
	\DontPrintSemicolon
	Initialization: $\pi_0$, $i=0$\;
	\While{$T^{\pi_i} \neq \emptyset$}{
		Select a non-empty and well-defined $U_i \subseteq \T[\pi_i]$\;
		$\pi_{i+1} = \pi_i \oplus U_i$\;
		$i \leftarrow i+1$\;
	}
	\Return $\pi_i$\;
\caption{Policy Iteration}
\end{algorithm} \label{algo:PI}

\begin{defi}[Policy Iteration]
	Algorithm~\ref{algo:PI} describes \emph{Policy Iteration} (PI). The standard way of choosing $U_i \subseteq \T[\pi_i]$ is the greedy update rule, namely choose any $U_i$ with maximal cardinality $|\S[\pi_i]|$. We refer to the corresponding algorithm as \emph{Greedy PI}, which is the focus of our work.
\end{defi}

\begin{defi}[Comparable]
	We say that two policies $\p$ and $\p'$ are \emph{comparable} if either $\p \preceq \p'$ or $\p \succeq \p'$. We call two policies \emph{neighbors} if they differ in only one state. Neighbors are always comparable (Lemma~3 in~\cite{Mansour1999}).
\end{defi}

\begin{defi}[Partial order]
	For a given MDP, we consider the \emph{partial order} PO of the policies defined by the domination relation. 
	A set of policies $\p^{(1)}, \dots, \p^{(k)}$ is called a \emph{sequence} if $\p^{(1)} \preceq \dots \preceq \p^{(k)}$. 
\end{defi}

\begin{defi}[PI-sequence]
	We refer to the sequence of policies $\p_0, \dots, \p_{m-1}$ explored by greedy PI as a \emph{PI-sequence} of length $m$. 
\end{defi}

We aim to solve the following problem.

\begin{pb} \label{def:problem}
	Find the longest possible PI-sequence.
\end{pb}

\begin{lem}[Lemma 4 in \cite{Mansour1999}] \label{thm:MS4}
	For any two policies $\p, \p'$ such that $\p'(s) = \p(s)$ for all improvement states $s \in \S$, we have $\p' \preceq \p$.
\end{lem}


The next property indicates how the improvement set of a policy is constrained by the dominated policies and by their own improvement sets.

\begin{prop} \label{thm:non-inclusion}
	For any two policies $\p \prec \p'$, there exists an improvement state $s \in \S$ such that $\p(s) \neq \p'(s)$ and $(s, \p(s)) \notin \T[\p']$.
	
	\begin{proof}
		Suppose on the contrary that it is not the case. Then for all states $s \in \S$, either $\p(s) = \p'(s)$ or $(s, \p(s)) \in \T[\p']$. Let $U \triangleq \big\{ \, (s, \p(s)) : s \in \S \cap \S[\p'] \text{ and } \p(s) \neq \p'(s) \, \big\}$, then we have $U \subseteq \T[\p']$. Therefore, Proposition~\ref{thm:MS1} tells us that $\p'' \triangleq \p' \oplus U \succeq \p'$. 
		
		Now, let us consider any $s \in \S$. If $\p'(s) = \p(s)$, then for any $a \in \Actions$, we have $(s, a) \notin U$ and $\p''(s) = \p(s)$. On the other hand, if $\p'(s) \neq \p(s)$, then $s \in \S[\p']$, hence $(s, \p(s)) \in U$ and $\p''(s) = \p(s)$ again. Therefore $\p''(s) = \p(s)$ for all $s \in \S$ and from Lemma~\ref{thm:MS4}, $\p'' \preceq \p \ (\prec \p')$ which is a contradiction.
	\end{proof}
\end{prop}

Note that for $k=2$, the statement of Proposition~\ref{thm:non-inclusion} can be simplified and implies that for any two policies $\p \prec \p'$, it holds that $\S \not\subseteq \S[\p']$.

When performing a PI step, we jump from the current policy to some policy that can be quite different (in terms of number of different entries). However, we now show that there always exists a path of small steps in the partial order connecting the two, that is, from neighbor to neighbor.

\begin{prop} \label{thm:k-path}
	Let $\p$ and $\p'$ be two policies such that $\p' = \p \oplus U$ for some well-defined $U \subseteq \T$ of cardinality $d$.
	Then there exist at least $d$ policies $\p^{(1)}, \dots, \p^{(d)}$ such that $\p \prec \p^{(1)} \preceq \dots \preceq \p^{(d)} = \p'$ and such that $\p^{(i)}$ and $\p^{(i+1)}$ are neighbors for all $1 \leq i < d$.
	
	\begin{proof}
		If $d = 1$, simply take $\p^{(d)} = \p'$. Suppose that the result is true for $d-1 \geq 1$ and let us show it for $d$. From Proposition~\ref{thm:non-inclusion}, there exists a state $s \in \S$ such that $(s, \p(s)) \notin \T[\p']$, that is, such that $\p' \oplus (s, \p(s)) \not\succ \p'$. Since neighbors are always comparable, it means that $\p'' \triangleq \p' \oplus (s, \p(s)) \preceq \p'$. By definition of $\p'$, we have $(s, \p'(s)) \in U$ and $U' \triangleq U \ \backslash \ (s, \p'(s)) \subseteq U \subseteq \T$. 
		We can then recursively apply the statement of Proposition~\ref{thm:k-path} with: 
		\begin{align*}
			\p' &\longmapsto \p'' = \pi' \oplus (s, \p(s)),\\
			U &\longmapsto U' = U \ \backslash \ (s, \p'(s)),
		\end{align*}
		since $\p'' = \p \oplus U'$ and $|U'| = d-1$.
		In that case, $\p^{(d-1)} = \p''$, and we can choose $\p^{(d)} = \p'$ which is indeed a neighbor of $\p^{(d-1)}$.
	\end{proof}
\end{prop}

\begin{defi}[Subsequence and supersequence]
	Let $O$ be a sequence. We call \emph{subsequence} of $O$ any ordered subset of elements of $O$. We call \emph{supersequence} of $O$ any sequence that contains $O$ as a subsequence.
\end{defi}

The following property is perhaps the most important consequence of Proposition~\ref{thm:k-path}.

\begin{cor}[Jumping] \label{thm:k-skip}
	Let $\p_i$ be a policy of a PI-sequence. Then the partial order of policies contains a supersequence of the PI-sequence with at least $|\S[\p_i]|$ different policies between $\p_i$ and $\p_{i+1}$, that is, $|\S[\p_i]|$ policies $\p$ such that $\p_i \prec \p \prec \p_{i+1}$.
	When we step from $\p_i$ to $\p_{i+1}$, we say that we \emph{jump} $|\S[\p_i]|$ policies of the supersequence.

	\begin{proof}
		The result is a direct consequence of Proposition~\ref{thm:k-path}. Recall that with Greedy PI, $|U_i|$ always equals $|\S[\p_i]|$. 
	\end{proof}
\end{cor}

\section{A relaxation of the problem} \label{sec:relaxation}


We now introduce an object that is similar to a PI-sequence in that it describes a sequence of policies embedded into a partial order. However, we will forget about some of the structure that originates from MDPs and only require Proposition~\ref{thm:non-inclusion} and Corollary~\ref{thm:k-skip} to be ensured by the sequence and the partial order.

\begin{defi}[Pseudo-PI-sequence] \label{def:pseudo-PI-seq}
	We call \emph{pseudo-PI-sequence} of size $m$ a triple $(\Pi, O, \mathcal{T})$ where:
	\begin{itemize}
		\item $\Pi = \p_0, \p_1, \dots, \p_{m-1}$ is a sequence of policies. We define the abstract ordering $\prec$ on the elements of the sequence $\Pi$ by the ordering of their indices.
		\item $O$ is a sequence of policies of $\{0,1\}^n$ that is a supersequence of $\Pi$.
		\item $\mathcal{T}$ is a collection of abstract improvement sets $\T$ for every policy $\p$ appearing in $O$.
	\end{itemize}
	We require the claim from Proposition~\ref{thm:non-inclusion} to hold for $O$ and we require $\Pi$ to satisfy Corollary~\ref{thm:k-skip} as a subsequence of $O$.
\end{defi}

Definition~\ref{def:pseudo-PI-seq} leads to a relaxation of Problem~\ref{def:problem}. Note that there is a natural way of constructing a pseudo-PI-sequence from any PI-sequence. Of course, Proposition~\ref{thm:non-inclusion} and Corollary~\ref{thm:k-skip}, that are the key results towards our upper bound in Theorem~\ref{thm:upper_bound}, still hold for pseudo-PI-sequences by design. Furthermore, as we will show in Theorem~\ref{thm:lower_bound}, our upper bound is tight for the relaxation.

\begin{rel} \label{def:relaxation}
	Find the longest possible pseudo-PI-sequence.
\end{rel}



In the rest of this paper, we only consider pseudo-PI-sequences and we now derive some of their properties. 
The following lemma is a direct consequence of Proposition~\ref{thm:non-inclusion}. 

\begin{lem} \label{thm:acyclicity}
	Let $(\Pi, O, \mathcal{T})$ be a pseudo-PI-sequence. Then for any two policies $\p \prec \p'$ of $O$ and any $U \subseteq \T[\p']$, we have $\p \neq \p' \oplus U$.

	\begin{proof}
		Let $s \in \S$ such that $\p'(s) \neq \p(s)$ and $(s, \p(s)) \notin \T[\p']$ whose existence is guaranteed by Proposition~\ref{thm:non-inclusion}. It is impossible to from switch $\p'(s)$ to $\p(s)$ hence the result.
	\end{proof}
\end{lem}


When $k = 2$, it is easy to see using Proposition~\ref{thm:non-inclusion} that two policies with exactly the same improvement states cannot exist. When $k > 2$, this is no longer the case. However, using Lemma~\ref{thm:MS4}, Mansour and Singh showed that there cannot be more than $k^d$ policies with the same $d$ improvement states in a PI-sequence (see Corollary~13 in~\cite{Mansour1999}). In the following proposition, we use Proposition~\ref{thm:non-inclusion} to improve this bound to $(k-1)^d$.

\begin{prop} \label{thm:k-1}
	Given a pseudo-PI-sequence $(\Pi, O, \mathcal{T})$ and a set of states $\S[] \subseteq \States$ of cardinality $d$, it holds that $O$ contains at most $(k-1)^d$ policies $\p$ with $\S = \S[]$.
	
	\begin{proof}
		Given the supersequence $O$ of the pseudo-PI-sequence, we consider its subsequence $\p^{(1)} \preceq \dots \preceq \p^{(K)}$ such that $\S[\p^{(i)}] = \S[] \triangleq \{ s_1, \dots, s_d \}$ for all $1 \leq i \leq K$. We show that if the subsequence satisfies Proposition~\ref{thm:non-inclusion}, then $K \leq (k-1)^d$. To this end, we first claim that the improvement sets of the policies of the subsequence can be assumed to be all well-defined. Indeed, for any policy of the subsequence $\p^{(i)}$, we can simplify its improvement set $\T[\p^{(i)}]$ by keeping only a single $(s, a)$ pair for every $s \in \S[\p^{(i)}]$. This does not modify $\S[\p^{(i)}]$ (i.e., $\p^{(i)}$ remains in the subsequence), nor does it imply the violation of Proposition~\ref{thm:non-inclusion}.
		Therefore, given a policy $\p^{(i)}$ of the subsequence and a state $s \in \S[]$, we can assume that there is exactly one action $a$ such that $(s, a) \in \T[\p^{(i)}]$, which we refer to as $\T[\p^{(i)}](s)$.
		
		We represent an action $i \in \Actions$ as a $k$-dimensional base vector $f_a(i) \triangleq e_i$ of $V = \mathbb{R}^k$,
		where $e_i(j) = 1$ if $i = j$, 0 otherwise. Similarly, we represent policies as base vectors of the space $W = V^{\otimes d}$ of dimension $k^d$ through the application:
		\begin{align*}
			f_p : \p \longmapsto f_a(\p(s_1)) \otimes \dots \otimes f_a(\p(s_d)).
		\end{align*}
		Finally, we represent pairs of policies and their improvement sets in a similar way in $W$ through the application:
		\begin{align*}
			f_c : (\p, \T) \longmapsto &\Big[ f_a(\p(s_1)) - f_a(\T(s_1)) \Big] \otimes \dots \otimes \Big[ f_a(\p(s_d)) - f_a(\T(s_d)) \Big],\\
			&= f_p(\p) \ \ + \sum_{ \begin{smallmatrix} U &\subseteq &\T \\ U &\neq &\emptyset \ \ \end{smallmatrix}} (-1)^{|U|} \cdot f_p(\p \oplus U).
		\end{align*}
		We claim that the vectors $f_c\big(\p^{(i)}, \T[\p^{(i)}]\big)$ are linearly independent. Assume on the contrary that we have:
		\begin{align}\label{eq:linearly-independent}%
			\sum_{i=1}^{K} \lambda_i \, f_c\left(\p^{(i)}, \T[\p^{(i)}]\right) = 0,
		\end{align}%
		with not all $\lambda_i$ being $0$. Choose the first index $i$ with non-zero $\lambda_i$. The corresponding term gives a non-zero coefficient to the base vector $f_p\big(\p^{(i)}\big)$. But from Lemma~\ref{thm:acyclicity}, for all $j > i$ and all $U \subseteq \T[\p^{(j)}]$, $\p^{(i)} \neq \p^{(j)} \oplus U$. Thus the base vector $f_p\big(\p^{(i)}\big)$ never appears later in the series in \eqref{eq:linearly-independent} which can therefore not be null.
		
		Additionally, the coordinates of $f_a(\p(s_i)) - f_a(\T(s_i)) \in V$ sum to 0 (in the standard base) for all $1 \leq i \leq d$ which means they lie in a subspace $V_0$ of $V$ of dimension $k-1$. As a result, 
		\begin{align*}
			f_c\left(\p^{(i)}, \T[\p^{(i)}] \right) \in W_0 = V_0^{\otimes d}.
		\end{align*}
		The dimension of $W_0$ is $(k-1)^d$ implying this is the maximum number of linearly independent vectors $f_c\big(\p^{(i)}, \T[\p^{(i)}] \big)$. This translates to the desired upper bound.
	\end{proof}
\end{prop}

Of course, the above result also holds for usual PI-sequences.

\section{Main result: a better upper bound on PI} \label{sec:thm1} \label{sec:upper-bound}

\begin{thm} \label{thm:upper_bound}
	The number of iterations of Policy Iteration is bounded above by 
	$\frac{k}{k-1} \cdot \frac{k^n}{n} + o\left( \frac{k^n}{n} \right)$.
	
	\begin{proof}
		The proof proceeds in two steps. First, we consider ``small'' improvement sets and show that there are at most $o\left(\frac{k^n}{n}\right)$ of them. Then we consider ``large'' improvement sets and show that PI explores at most $\frac{k}{k-1} \cdot \frac{k^n}{n} + o\left( \frac{k^n}{n} \right)$ of them because they jump many policies on the way.
		
		\textbf{Small improvement sets.} We consider the small improvement sets $\T$ such that $|\S| \leq \frac{k-1}{k} \cdot n - f(n)$ with:
		\begin{align*}
			f(n) \triangleq \sqrt{n \log n}.
		\end{align*}
		From Proposition~\ref{thm:k-1}, policies with the same set of improvement states $S$ of cardinality $d$ can appear at most $(k-1)^d$ times in a (pseudo-)PI-sequence, hence the number of small improvement sets can be expressed as follows:
		\begin{align*}
			\sum_{d = 0}^{\floor*{\frac{k-1}{k} \cdot n - f(n)}} \nchoosek{n}{d} (k-1)^d
			& = k^n \hspace{-.2cm} \sum_{d = 0}^{\floor*{\frac{k-1}{k} \cdot n - f(n)}} \hspace{-.1cm} \nchoosek{n}{d} \left(\frac{k-1}{k}\right)^{\hspace{-.05cm} d} \hspace{-.05cm} \left(\frac{1}{k}\right)^{\hspace{-.05cm} n-d}\hspace{-.5cm}, \\
			& = k^n \cdot P\left[ X \leq \frac{k-1}{k} \cdot n - f(n) \right], 
		\end{align*}
		where $X \sim \mathrm{Bin}\left(n, \frac{k-1}{k}\right)$ follows a binomial distribution. 
		Using Hoeffding's inequality~\cite{hoeffding1963}, we have:
		\begin{align*}
			P\left[ X \leq n \cdot \left(\frac{k-1}{k} - \frac{f(n)}{n}\right) \right] 
			&\leq e^{-2 \cdot \left(\frac{f(n)}{n}\right)^2 \cdot n}
			= \frac{1}{n^2}.
		\end{align*}
		Therefore we have:
		\begin{align*}
			\sum_{d = 0}^{\floor*{\frac{k-1}{k} \cdot n - f(n)}} \nchoosek{n}{d} (k-1)^d 
			& \leq k^n \cdot \frac{1}{n^2} = o\left(\frac{k^n}{n}\right).
		\end{align*}
		
		\textbf{Large improvement sets.} We now consider the improvement sets $\T$ with the set of improvement states satisfying $|\S| > \frac{k-1}{k} \cdot n - f(n)$. We show that these sets jump many policies on the way and hence we cannot have many of them in the (pseudo-)PI-sequence. Suppose that we have $K$ such improvement sets in the sequence. Then, from Corollary~\ref{thm:k-skip}, we jump at least $K \cdot \big(\frac{k-1}{k} \cdot n - f(n)\big)$ distinct policies. Since we cannot jump more that $k^n$ policies, we have the following condition on $K$:
		\begin{align*}
			K \leq \frac{k^n}{\frac{k-1}{k} n - f(n)} = \frac{k}{k-1} \cdot \frac{k^n}{n} \cdot \frac{1}{1-\frac{k-1}{k} \sqrt{\frac{\log n}{n}}}.
		\end{align*}
		Hence $K~\leq~\frac{k}{k-1}~\cdot~\frac{k^n}{n} + o\left( \frac{k^n}{n} \right)$.
	\end{proof}
	
\end{thm}


\section{The bound is tight for the relaxation} \label{sec:thm2} \label{sec:lower-bound}


\begin{SCfigure}[2.3]
\caption{An example of a pseudo-PI-Sequence of size $\frac{k}{k-1}~\cdot~\frac{k^n}{n} + o\left(\frac{k^n}{n}\right)$ with its supersequence $O$ for $n = k = 3$. Each gray box corresponds to a policy of the supersequence. We represent the improvement sets only through the prospective improving action for each state (action $3$ for state $s$ if $\p(s) \neq 3$ or nothing, according to the construction). The red policies are the ones from the sequence $\Pi$ from Definition~\ref{def:pseudo-PI-seq}. It can be checked that if some policy $\p_i$ is in $\mathcal{T}^d$, then $d$ policies of the supersequence are jumped from $\p_i$ to $\p_{i+1}$ and it can be observed that the supersequence contains $k^n$ elements and satisfies the claim of Proposition~\ref{thm:non-inclusion}.}
\scalebox{.75}{%
\begin{tikzpicture}
	\xdef\smalldy{.25}
	\xdef\largedy{1.25}
	\xdef\rectx{.25}
	\xdef\recty{1.75}
	\xdef\xp{-.3}
	\xdef\xa{-1.7}
	\xdef\xb{2.8}
	
	\edef\pos{0}
	\foreach \d in {3, 2, 1, 0}
	{
		\xdef\factoriald{1}
		\pgfmathparse{max(1,\d)}
		\foreach \i in {1, ..., \pgfmathresult}
		{
			\pgfmathparse{\factoriald*\i}
			\xdef\factoriald{\pgfmathresult}
		}
		\xdef\factorialnd{1}
		\pgfmathparse{max(1,3-\d)}
		\foreach \i in {1, ..., \pgfmathresult}
		{
			\pgfmathparse{\factorialnd*\i}
			\xdef\factorialnd{\pgfmathresult}
		}
		\pgfmathparse{6/\factoriald/\factorialnd}
		\xdef\nck{\pgfmathresult}
		\pgfmathparse{\nck-1}
		\xdef\nckm{\pgfmathresult}
		
		\foreach \b in {0, ..., \nckm}
		{
			\pgfmathparse{pow(2,\d)}
			\xdef\pow{\pgfmathresult}
			\pgfmathparse{\pow-1}
			\xdef\powm{\pgfmathresult}
			\foreach \a in {0, ..., \powm}
			{
				\pgfmathparse{\pos+1}
				\xdef\pos{\pgfmathresult}
				\pgfmathparse{\pow*\b +\a}
				\xdef\xpos{\pgfmathresult}
						
				\xdef\entryI{3}
				\ifthenelse{\equal{\d}{3} \AND \equal{\b}{0} \AND \a < 4}{\xdef\entryI{1}}{}
				\ifthenelse{\equal{\d}{2} \AND \b < 2 		 \AND \a < 2}{\xdef\entryI{1}}{}
				\ifthenelse{\equal{\d}{1} \AND \equal{\b}{0} \AND \a < 1}{\xdef\entryI{1}}{}
				\ifthenelse{\equal{\d}{3} \AND \equal{\b}{0} \AND \a > 3}{\xdef\entryI{2}}{}
				\ifthenelse{\equal{\d}{2} \AND \b < 2 		 \AND \a > 1}{\xdef\entryI{2}}{}
				\ifthenelse{\equal{\d}{1} \AND \equal{\b}{0} \AND \a > 0}{\xdef\entryI{2}}{}
				
				\xdef\entryII{3}
				\ifthenelse{\equal{\d}{3} \AND \equal{\b}{0} \AND \a < 2            }{\xdef\entryII{1}}{}
				\ifthenelse{\equal{\d}{3} \AND \equal{\b}{0} \AND \a < 6 \AND \a > 3}{\xdef\entryII{1}}{}
				\ifthenelse{\equal{\d}{2} \AND \equal{\b}{0} \AND \equal{\a}{0}     }{\xdef\entryII{1}}{}
				\ifthenelse{\equal{\d}{2} \AND \equal{\b}{0} \AND \equal{\a}{2}     }{\xdef\entryII{1}}{}
				\ifthenelse{\equal{\d}{2} \AND \equal{\b}{2} \AND \equal{\a}{0}     }{\xdef\entryII{1}}{}
				\ifthenelse{\equal{\d}{2} \AND \equal{\b}{2} \AND \equal{\a}{1}     }{\xdef\entryII{1}}{}
				\ifthenelse{\equal{\d}{1} \AND \equal{\b}{1} \AND \a < 1			}{\xdef\entryII{1}}{}
				\ifthenelse{\equal{\d}{3} \AND \equal{\b}{0} \AND \a < 4 \AND \a > 1}{\xdef\entryII{2}}{}
				\ifthenelse{\equal{\d}{3} \AND \equal{\b}{0} \AND \a < 8 \AND \a > 5}{\xdef\entryII{2}}{}
				\ifthenelse{\equal{\d}{2} \AND \equal{\b}{0} \AND \equal{\a}{1}     }{\xdef\entryII{2}}{}
				\ifthenelse{\equal{\d}{2} \AND \equal{\b}{0} \AND \equal{\a}{3}     }{\xdef\entryII{2}}{}
				\ifthenelse{\equal{\d}{2} \AND \equal{\b}{2} \AND \equal{\a}{2}     }{\xdef\entryII{2}}{}
				\ifthenelse{\equal{\d}{2} \AND \equal{\b}{2} \AND \equal{\a}{3}     }{\xdef\entryII{2}}{}
				\ifthenelse{\equal{\d}{1} \AND \equal{\b}{1} \AND \a > 0			}{\xdef\entryII{2}}{}
				
				\xdef\entryIII{3}
				\ifthenelse{\equal{\d}{3} \AND \equal{\b}{0} \AND \equal{\a}{0}}{\xdef\entryIII{1}}{}
				\ifthenelse{\equal{\d}{3} \AND \equal{\b}{0} \AND \equal{\a}{2}}{\xdef\entryIII{1}}{}
				\ifthenelse{\equal{\d}{3} \AND \equal{\b}{0} \AND \equal{\a}{4}}{\xdef\entryIII{1}}{}
				\ifthenelse{\equal{\d}{3} \AND \equal{\b}{0} \AND \equal{\a}{6}}{\xdef\entryIII{1}}{}
				\ifthenelse{\equal{\d}{2} \AND \b > 0 		 \AND \equal{\a}{0}}{\xdef\entryIII{1}}{}
				\ifthenelse{\equal{\d}{2} \AND \b > 0 		 \AND \equal{\a}{2}}{\xdef\entryIII{1}}{}
				\ifthenelse{\equal{\d}{1} \AND \equal{\b}{2} \AND \equal{\a}{0}}{\xdef\entryIII{1}}{}
				\ifthenelse{\equal{\d}{3} \AND \equal{\b}{0} \AND \equal{\a}{1}}{\xdef\entryIII{2}}{}
				\ifthenelse{\equal{\d}{3} \AND \equal{\b}{0} \AND \equal{\a}{3}}{\xdef\entryIII{2}}{}
				\ifthenelse{\equal{\d}{3} \AND \equal{\b}{0} \AND \equal{\a}{5}}{\xdef\entryIII{2}}{}
				\ifthenelse{\equal{\d}{3} \AND \equal{\b}{0} \AND \equal{\a}{7}}{\xdef\entryIII{2}}{}
				\ifthenelse{\equal{\d}{2} \AND \b > 0 		 \AND \equal{\a}{1}}{\xdef\entryIII{2}}{}
				\ifthenelse{\equal{\d}{2} \AND \b > 0 		 \AND \equal{\a}{3}}{\xdef\entryIII{2}}{}
				\ifthenelse{\equal{\d}{1} \AND \equal{\b}{2} \AND \equal{\a}{1}}{\xdef\entryIII{2}}{}
				
				\xdef\sI{0}
				\ifthenelse{\equal{\d}{3}			 }{\xdef\sI{1}}{}
				\ifthenelse{\equal{\d}{2} \AND \b < 2}{\xdef\sI{1}}{}
				\ifthenelse{\equal{\d}{1} \AND \b < 1}{\xdef\sI{1}}{}
				\xdef\sII{0}
				\ifthenelse{\equal{\d}{3}			 }{\xdef\sII{1}}{}
				\ifthenelse{\equal{\d}{2} \AND \b < 1}{\xdef\sII{1}}{}
				\ifthenelse{\equal{\d}{2} \AND \b > 1}{\xdef\sII{1}}{}
				\ifthenelse{\equal{\d}{1} \AND \equal{\b}{1}}{\xdef\sII{1}}{}
				\xdef\sIII{0}
				\ifthenelse{\equal{\d}{3}			 }{\xdef\sIII{1}}{}
				\ifthenelse{\equal{\d}{2} \AND \b > 0}{\xdef\sIII{1}}{}
				\ifthenelse{\equal{\d}{1} \AND \b > 1}{\xdef\sIII{1}}{}
				
				\pgfmathparse{0}
				\xdef\xI{\pgfmathresult}
				\pgfmathparse{1}
				\xdef\xII{\pgfmathresult}
				\pgfmathparse{2}
				\xdef\xIII{\pgfmathresult}
				\pgfmathparse{-\pos*\largedy+\smalldy}
				\xdef\yI{\pgfmathresult}
				\pgfmathparse{-\pos*\largedy		 }
				\xdef\yII{\pgfmathresult}
				\pgfmathparse{-\pos*\largedy-\smalldy}
				\xdef\yIII{\pgfmathresult}
				
				\fill[black!5!white] (\xI-\rectx, \yII-.45) rectangle (\xIII+\rectx, \yII+.45);
				
				\draw (\xII , \yII ) node {{\footnotesize $\ra$}};
				
				\ifthenelse{\equal{\sI}{1}}
				{
					\ifthenelse{\equal{\pos}{1.0} \OR \equal{\pos}{4.0} \OR \equal{\pos}{7.0} \OR \equal{\pos}{10.0} \OR \equal{\pos}{12.0} \OR \equal{\pos}{14.0} \OR \equal{\pos}{16.0} \OR \equal{\pos}{18.0} \OR \equal{\pos}{20.0} \OR \equal{\pos}{22.0} \OR \equal{\pos}{23.0} \OR \equal{\pos}{24.0} \OR \equal{\pos}{25.0} \OR \equal{\pos}{26.0}}
					{
						\draw (\xI  , \yI  ) node {{\footnotesize $\color{red!30!green}{\entryI}$}};
					}{
						\draw (\xI  , \yI  ) node {{\footnotesize $\entryI$}};
					}
					\draw (\xIII, \yI  ) node {{\footnotesize $3$}};
				}{
					\draw (\xI  , \yI  ) node {{\footnotesize $\entryI$}};
				}
				\ifthenelse{\equal{\sII}{1}}
				{
					\ifthenelse{\equal{\pos}{1.0} \OR \equal{\pos}{4.0} \OR \equal{\pos}{7.0} \OR \equal{\pos}{10.0} \OR \equal{\pos}{12.0} \OR \equal{\pos}{14.0} \OR \equal{\pos}{16.0} \OR \equal{\pos}{18.0} \OR \equal{\pos}{20.0} \OR \equal{\pos}{22.0} \OR \equal{\pos}{23.0} \OR \equal{\pos}{24.0} \OR \equal{\pos}{25.0} \OR \equal{\pos}{26.0}}
					{
						\draw (\xI  , \yII ) node {{\footnotesize $\color{red!30!green}{\entryII}$}};
					}{
						\draw (\xI  , \yII ) node {{\footnotesize $\entryII$}};
					}
					\draw (\xIII, \yII ) node {{\footnotesize $3$}};
				}{
					\draw (\xI  , \yII ) node {{\footnotesize $\entryII$}};
				}
				\ifthenelse{\equal{\sIII}{1}}
				{
					\ifthenelse{\equal{\pos}{1.0} \OR \equal{\pos}{4.0} \OR \equal{\pos}{7.0} \OR \equal{\pos}{10.0} \OR \equal{\pos}{12.0} \OR \equal{\pos}{14.0} \OR \equal{\pos}{16.0} \OR \equal{\pos}{18.0} \OR \equal{\pos}{20.0} \OR \equal{\pos}{22.0} \OR \equal{\pos}{23.0} \OR \equal{\pos}{24.0} \OR \equal{\pos}{25.0} \OR \equal{\pos}{26.0}}
					{
						\draw (\xI  , \yIII) node {{\footnotesize $\color{red!30!green}{\entryIII}$}};
					}{
						\draw (\xI  , \yIII) node {{\footnotesize $\entryIII$}};
					}
					\draw (\xIII, \yIII) node {{\footnotesize $3$}};
				}{
					\draw (\xI  , \yIII) node {{\footnotesize $\entryIII$}};
				}
			}
		}
	}
	
	\node[draw=none] at (\xI  , 0) {$\p$};
	\node[draw=none] at (\xIII, 0) {$\T$};
	\draw (-1.2,-.4) -- (2.5,-.4);
		
	\foreach \i in {0, 1, 2}
	{
		\pgfmathparse{-3*\i*\largedy-\largedy}
		\xdef\yc{\pgfmathresult}
		\node[draw=none,anchor=east] at (\xp, \yc) {{$\mathbf{\color{red!70!green}{\pi_{\i}=}}$}};
		\draw[->,line width=2pt,color=red!70!green,shorten >=3pt] (\xa, \yc) to[bend right] (\xa, \yc-3*\largedy);
		\draw[color=red!70!green] (\xI-\rectx,\yc-\recty*\smalldy) rectangle (\xI+\rectx,\yc+\recty*\smalldy);
	}
	\foreach \i in {3, 4, 5, 6, 7, 8}
	{
		\pgfmathparse{-2*\i*\largedy-4*\largedy}
		\xdef\yc{\pgfmathresult}
		\node[draw=none,anchor=east] at (\xp, \yc) {{$\mathbf{\color{red!70!green}{\pi_{\i}=}}$}};
		\draw[->,line width=2pt,color=red!70!green,shorten >=3pt] (\xa, \yc) to[bend right] (\xa, \yc-2*\largedy);
		\draw[color=red!70!green] (\xI-\rectx,\yc-\recty*\smalldy) rectangle (\xI+\rectx,\yc+\recty*\smalldy);
	}
	\foreach \i in {9, 10, 11, 12, 13}
	{
		\pgfmathparse{-\i*\largedy-13*\largedy}
		\xdef\yc{\pgfmathresult}
		\node[draw=none,anchor=east] at (\xp, \yc) {{$\mathbf{\color{red!70!green}{\pi_{\i}=}}$}};
		\draw[->,color=red!70!green,line width=2pt,shorten >=3pt] (\xa, \yc) to[bend right] (\xa, \yc-\largedy);
		\draw[color=red!70!green] (\xI-\rectx,\yc-\recty*\smalldy) rectangle (\xI+\rectx,\yc+\recty*\smalldy);
	}
	\pgfmathparse{-27*\largedy}
	\xdef\yc{\pgfmathresult}
	\node[draw=none,anchor=east] at (\xp, \yc) {{$\mathbf{\color{red!70!green}{\pi_{14}=}}$}};
	\draw[color=red!70!green] (\xI-\rectx,\yc-\recty*\smalldy) rectangle (\xI+\rectx,\yc+\recty*\smalldy);

	\foreach \i in {1, ..., 26}
	{
		\pgfmathparse{-\i*\largedy-.5*\largedy}
		\xdef\yc{\pgfmathresult}
		\node[draw=none,rotate=90] at (\xI, \yc) {{\scriptsize $\succ$}};
	}

	\draw [decorate,decoration={brace,amplitude=8pt},yshift=0pt,red!30!green,line width=2pt] (\xb,-\largedy+\recty*\smalldy) -- (\xb,-8*\largedy-\recty*\smalldy) node [midway,right,xshift=9pt] {$\mathbf{\color{red!30!green}\mathcal{T}^3}$};
	\draw [decorate,decoration={brace,amplitude=8pt},yshift=0pt,red!30!green,line width=2pt] (\xb,-9*\largedy+\recty*\smalldy) -- (\xb,-20*\largedy-\recty*\smalldy) node [midway,right,xshift=9pt] {$\mathbf{\color{red!30!green}\mathcal{T}^2}$};
	\draw [decorate,decoration={brace,amplitude=8pt},yshift=0pt,red!30!green,line width=2pt] (\xb,-21*\largedy+\recty*\smalldy) -- (\xb,-26*\largedy-\recty*\smalldy) node [midway,right,xshift=9pt] {$\mathbf{\color{red!30!green}\mathcal{T}^1}$};
	\draw [decorate,decoration={brace,amplitude=5pt},yshift=0pt,red!30!green,line width=2pt] (\xb,-27*\largedy+\recty*\smalldy) -- (\xb,-27*\largedy-\recty*\smalldy) node [midway,right,xshift=9pt] {$\mathbf{\color{red!30!green}\mathcal{T}^0}$};
	
\end{tikzpicture}
}
\end{SCfigure}
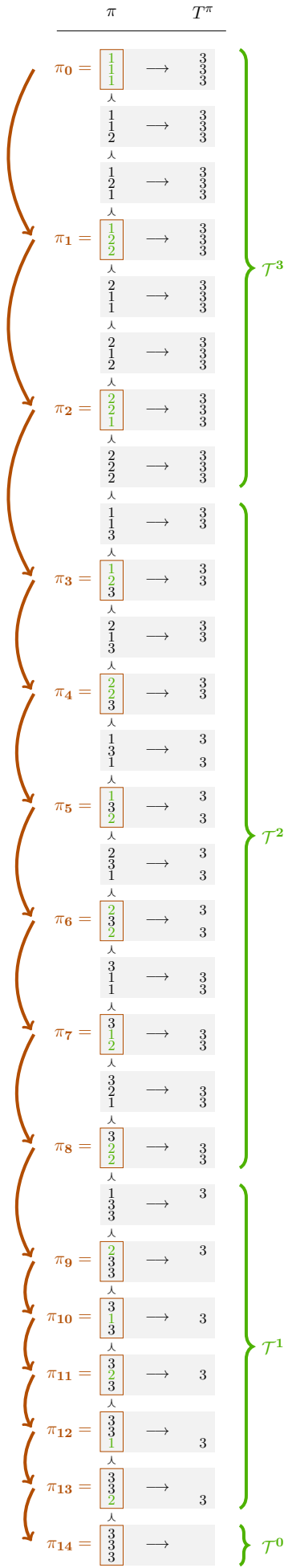

The following theorem shows that the upper bound from Theorem~\ref{thm:upper_bound} is tight for Relaxation~\ref{def:relaxation}.

\begin{thm} \label{thm:lower_bound}
	There exists a pseudo-PI-sequence of size $\frac{k}{k-1} \cdot \frac{k^n}{n} + o\left( \frac{k^n}{n} \right)$.
	
	\begin{proof}
		We first build a sequence containing all the $k^n$ policies that will play the role of the supersequence $O$ for the pseudo-PI-sequence. Preliminarily, given any policy $\p$ of $O$, we define its (well-defined) improvement set $\T$ such that $(s, a) \in \T$ iff $\p(s) \neq k$ and $a = k$. Here action $k$ can be thought of as some special action. Let $\mathcal{T}^d$ be the set of all policies $\p$ such that $|\T| = d$. By definition, $\mathcal{T}^d$ contains 
		all policies $\p$ such that $\p(s) \neq k$ for exactly $d$ different states $s$, hence 
		$\smallnchoosek{n}{d}\cdot(k-1)^d$ elements. We now order all $k^n$ policies as a sequence by decreasing order of cardinality of their improvement sets, hence the policies in $\mathcal{T}^d$-sets with a large $d$ come first in the sequence. The (total) ordering inside a given $\mathcal{T}^d$-set can be arbitrarily chosen. Given this ordering, notice that for any $\p \prec \p'$, if $\S \subseteq \S[\p']$, then $\S = \S[\p']$.
		

		
		The sequence $O$ obtained with the above construction satisfies the claim of Proposition~\ref{thm:non-inclusion}. Indeed, let us choose any two policies of the sequence $\p \prec \p'$. First assume that $\S \ \backslash \ \S[\p'] \neq \emptyset$ and let $t \in \S \ \backslash \ \S[\p']$. Then by construction, $\p(t) \neq k = \p'(t)$ and $(t, \p(t)) \notin \T[\p']$ since $t \notin \S[\p']$, hence Proposition~\ref{thm:non-inclusion} is true in that case. If now $\S \ \backslash \ \S[\p'] = \emptyset$, then the ordering of the policies imposes that $\S = \S[\p']$, as observed above. In that case, by construction $\p(s) \neq k$ for all $s \in \S$ and $\p(s) = \p'(s) = k$ for all $s \notin \S$. Since $\p \neq \p'$, there must exist some state $t \in \S$ such that $\p(t) \neq \p'(t)$. Furthermore by definition of $\T[\p']$, $(t, \p(t)) \notin \T[\p']$ because $\p(t) \neq k$, and the claim of Proposition~\ref{thm:non-inclusion} is true again.
		

		At this point, we have built a supersequence for our PI-sequence that satisfies the claim of Proposition~\ref{thm:non-inclusion}. Let us now select a subsequence $\Pi$ of $O$ while ensuring Corollary~\ref{thm:k-skip} as follows: we start from the first policy of the supersequence $\p_0$, $i=0$. Then at each step $i$, we jump $|\T[\p_i]|$ elements in the sequence to select $\p_{i+1}$. With this greedy procedure, we clearly ensure Corollary~\ref{thm:k-skip} and we pick at least $\frac{1}{d+1} |\mathcal{T}^d|$ policies from each $\mathcal{T}^d$-set, for a total number of hypothetical PI-steps of at least:
		\begin{align*}
			& \sum_{d=0}^{n} \frac{1}{d+1} |\mathcal{T}^d|,\\
			& = \sum_{d=0}^{n} \frac{1}{d+1} \nchoosek{n}{d} (k-1)^d,\\
			& = \frac{1}{n+1} \cdot \sum_{d=0}^{n} \nchoosek{n+1}{d+1} \cdot (k-1)^{d} \cdot 1^{n - d},\\
			& = \frac{1}{k-1} \cdot \frac{1}{n+1} \cdot \Bigg[ \underbrace{\sum_{d=0}^{n+1} \nchoosek{n+1}{d} \cdot (k-1)^d \cdot 1^{(n+1) - d}}_{= k^{n+1}} \ - \ 1 \Bigg],\\
			& = \frac{k}{k-1} \cdot \frac{k^n}{n} + o\left( \frac{k^n}{n} \right),
		\end{align*} 
		which corresponds to our claim and matches the upper bound from Theorem~\ref{thm:upper_bound}. An example of a pseudo-PI-sequence constructed from the above procedure with $n = k = 3$ is given in Figure~\ref{fig:order_regular_examples}.
	\end{proof}
	
\end{thm}

Of course, the lower bound from Theorem~\ref{thm:lower_bound} only holds for pseudo-PI-sequences which are much less constrained than usual PI-sequences. Indeed, it can be observed that the pseudo-PI-sequence constructed above cannot correspond to a real PI-run since for instance its supersequence does not satisfy Proposition~\ref{thm:MS1}. Therefore, obtaining better bounds than the one from Theorem~\ref{thm:upper_bound} will require a more advanced analysis as discussed in the next section.

\section{Alternative approaches} \label{sec:perspectives}

Theorem~\ref{thm:lower_bound} revealed that future improvements of our bound will require to take into account more of the combinatorial structure of PI-sequences. In this section, we describe two advanced approaches that could lead to new results.

The idea of the first approach is to represent the partial order of the policies of an MDP as an oriented graph whose nodes---the policies---are embedded in an $n$-dimensional grid and whose edges---that translate the domination relation---only connect neighboring policies (that is, that differ in only one state). In this framework, the structure of the partial order is best described by the Acyclic Unique Sink Orientation of a Grid\footnote{One could strengthen even a bit further by requiring the Holt-Klee condition as well \cite{holt1999,gartner2005}.} (Grid AUSOs), an object introduced by G\"artner et al~\cite{gartner2005} as a generalization of Acyclic Unique Sink Orientations of Cubes~\cite{szabo2001} when $k > 2$.
Grid AUSOs accurately characterize the structure of the partial order of any MDP and essentially all necessary conditions we know on PI-sequences originate from this framework. More precisely, it can be described as follows: take a Cartesian grid of dimension $n$, the number of states of the MDP. A policy can be represented by its action at each state as a vector in $\{1, ..., k\}^n$ and it thereby corresponds to a vertex of the grid. 
For every neighboring policies $\pi, \pi'$, we draw a directed edge from $\pi$ to $\pi'$ if $\pi \prec \pi'$ (recall that neighboring policies are always comparable). Thereby, we obtain a directed graph on the grid that is guaranteed to be acyclic and unique sink, i.e. any sub-grid of dimension $d \leq n$ contains a unique vertex of maximum in-degree $d$~\cite{gartner2005}.

With this structure, PI-steps can be viewed as jumps in the grid as follows: from a policy $\pi_i$ of the PI-sequence, the out-going links at the corresponding vertex span a sub-grid. In general, the next vertex $\pi_{i+1}$ chosen by PI can be any vertex of this sub-grid, but in the greedy version, some antipodal vertex to $\pi_i$ is chosen. This algorithm is also known as the Bottom-Antipodal method in the AUSO framework. Note that it is possible to design Cube AUSOs for which PI takes $\Omega\big(\sqrt{2}^n\big)$ steps~\cite{schurr2005} but to the best of our knowledge, this lower bound cannot be adapted for MDPs.


For $k = 2$, another promising approach was proposed by Hansen \& Zwick~\cite{Hansen2012} through a relaxation of the AUSO structure. Their idea is to record the policies visited by PI in a binary matrix $\Pi \in \{0,1\}^{m \times n}$ whose columns correspond to the states of the MDP and whose $(i+1)^{\text{th}}$ row corresponds to the policy $\pi_i$ of a PI-sequence. They then formulate the following combinatorial condition on this matrix: for every rows $i,j$ of $\Pi, i<j$, there must exist a column $k$ such that:
\begin{align} \label{eq:matrix_condition}
	\Pi_{i,k} \neq \Pi_{i+1,k} = \Pi_{j,k} = \Pi_{j+1,k}.
\end{align}
In case $j+1 > m$, we use the convention that $\Pi_{m+1,k} = \Pi_{m,k}$. Furthermore, the last two rows (labeled $m-1$ and $m$) are required to be distinct. Intuitively, 
at each step $i < m$, at least one change is made to the policy (otherwise we have convergence). Then, at any later step $j$, one of the changes made at step $i$ must still be there and accepted for the next step. An upper bound on the number of rows of such matrices would immediately translate in a bound on the length of PI-sequences.

\begin{table}[h!]
  \centering
  {\tt
  \begin{tabular}{ccc}
    0&0&0\\
    1&1&1\\
    0&0&1\\
    0&1&1\\
    0&1&0
  \end{tabular}
  \hspace{2cm}
  \begin{tabular}{cccc}
    0&0&0&0\\
    1&1&1&1\\
    0&0&0&1\\
    0&1&1&1\\
    0&0&1&0\\
    0&1&1&0\\
    0&1&0&0\\
    1&1&0&0
  \end{tabular}}
  \caption{Examples of extremal matrices satisfying condition \eqref{eq:matrix_condition} for 3 and 4 columns.}
  \label{fig:order_regular_examples}
\end{table}

Simulations hint towards the Fibonacci sequence as a possible upper bound for this relaxed problem: for $n \leq 6$, extremal instances achieve $m = F_{n+2}$, the $(n+2)^{\text{nd}}$ Fibonacci number. 
If true, this bound would be a significant improvement to ours in the case where $k=2$. Note that it is possible to build matrices with $m = \Omega\big(\sqrt{2}^n\big)$ rows using similar constructions as for AUSOs. Improving these lower bounds is an interesting challenge in itself.

\section{Summary}

Our contributions can be summarized as follows. First in Theorem~\ref{thm:upper_bound}, we show that Policy Iteration cannot take more than $\frac{k}{k-1} \cdot \frac{k^n}{n} + o\big(\frac{k^n}{n}\big)$ steps to converge, independently of the chosen reward criterion for the MDP. We thereby improve Mansour and Singh's fifteen years old bound. Then in Theorem~\ref{thm:lower_bound}, we show that our bound is optimal for some natural relaxation of the problem. Finally in Section~\ref{sec:perspectives}, we survey two advanced combinatorial approaches that still could lead to an improvement to our bound.


\bibliographystyle{alpha}
\bibliography{biblio}

\end{document}